\documentclass[aps,prb,twocolumn,groupedaddress,floatfix,showpacs]{revtex4}

\usepackage{graphicx}
\usepackage{dcolumn} 
\usepackage{bm}      
\usepackage{amsmath} 

\begin{document}

\title{Charge order in Fe$_2$OBO$_3$: An LSDA+$U$ study}

\author{I.~Leonov$^{1}$\email[e-mail: ]{Ivan.Leonov@Physik.uni-Augsburg.de},
A.~N.~Yaresko$^2$, V.~N.~Antonov$^3$, J.~P.~Attfield$^4$, and V.~I.~Anisimov$^5$}

\affiliation{$^1$ Theoretical Physics III, Center for Electronic Correlations and Magnetism, 
Institute for Physics, University of Augsburg, Germany}
\affiliation{$^2$ Max-Planck Institute for the Physics of Complex Systems, Dresden, Germany}
\affiliation{$^3$ Institute of Metal Physics, Vernadskii Street, 03142 Kiev, Ukraine}
\affiliation{$^4$ Centre for Science at Extreme Conditions, University of Edinburgh, 
Erskine, Williamson Building, King`s Buildings, Mayfield Road, Edinburgh EH9 3JZ, United Kingdom}
\affiliation{$^5$ Institute of Metal Physics, Russian Academy of Science-Ural Division, 620219 
Yekaterinburg GSP-170, Russia}

\date{\today}

\begin{abstract}
Charge ordering in the low-temperature monoclinic structure of iron oxoborate
(Fe$_2$OBO$_3$) is investigated using the local spin density approximation 
(LSDA)+$U$ method. While the difference between $t_{2g}$ minority occupancies 
of Fe$^{2+}$ and Fe$^{3+}$ cations is large and gives direct evidence for 
charge ordering, the static ``screening'' is so effective that the total $3d$ 
charge separation is rather small. The occupied Fe$^{2+}$ and Fe$^{3+}$ 
cations are ordered alternately within the chain which is infinite along the 
$a$-direction. The charge order obtained by LSDA+$U$ is consistent with 
observed enlargement of the $\beta$ angle. An analysis of the exchange 
interaction parameters demonstrates the predominance of the interribbon 
exchange interactions which determine the whole $L$-type ferrimagnetic 
spin structure. 

\end{abstract}


\maketitle


\section{Introduction}
\label{sec:introd}

Transition metal compounds with interplay of spin, orbital, and charge degrees 
of freedom are of strong current interest. Thus, magnetite (Fe$_3$O$_4$), the 
famous lodestone,\cite{Mattis88} is one of the classical examples of such a 
system, where a first-order metal-insulator transition occurs at 
$\sim 120$ K.\cite{Rev01} According to Verwey this transition is caused by 
the ordering of Fe$^{2+}$ and Fe$^{3+}$ cations on the octahedral $B$-sublattice 
of the inverted spinel structure AB$_2$O$_4$.\cite{V39,VHR47} Recently, the 
local spin density approximation (LSDA)+$U$ method~\cite{AZA91,LAZ95,AAL97} 
has been used for investigation of the low-temperature monoclinic 
structure~\cite{WAR01,WAR02} of Fe$_3$O$_4$~\cite{LYA04}. The charge 
and orbitally ordered insulating ground state with two different order parameters: 
small for the charge and large for the orbital order was obtained self-consistently. 
While the first order parameter was introduced as the total $3d$ charge separation 
between Fe$^{2+}$ and Fe$^{3+}$ cations, the latter one corresponds to the charge 
difference between occupancies of occupied $t_{2g\downarrow}$ orbital of Fe$^{2+}$ 
and unoccupied $t_{2g\downarrow}$ orbital of Fe$^{3+}$ cations. But an important 
issue, which is still unresolved is whether the charge ordering in magnetite
is driven by Coulomb repulsion between the charges, or by the strain arising 
from electron-lattice interactions. An analysis of the charge and orbital order 
parameters~\cite{LYA04} in Fe$_3$O$_4$ inevitably results in a strong interplay 
between Jahn-Teller effect and electrostatic repulsion between electrons. Therefore, 
theoretical investigations of the isostructural iron and manganese oxoborate 
(Fe$_2$OBO$_3$ and Mn$_2$OBO$_3$, correspondingly) that support predominantly 
electrostatic and Jahn-Teller distortion driven mechanisms~\cite{ABRM98} behind 
the charge ordering in the same structural arrangement, are highly desirable. 

\begin{figure}[tbp!]
\centerline{\includegraphics[width=0.25\textwidth,clip]{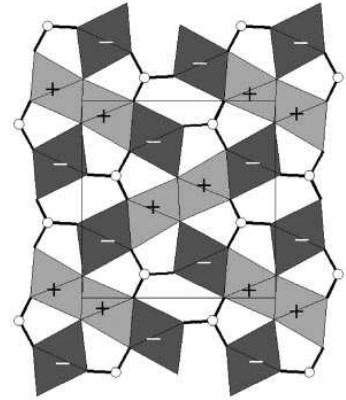}}
\caption{\label{fig:struc}
The Fe$_2$OBO$_3$ structure projected on (100) plane ($b$ vertical, $c$ horizontal).
Two structurally distinct Fe(1)O$_6$ and Fe(2)O$_6$ octahedra shown by light and 
dark sharing respectively. Plus and minus signs indicate the relative orientation
of the moments within each Fe(1) and Fe(2) chains in the magnetically ordered phase. }
\end{figure}

Iron borate (Fe$_2$OBO$_3$) is a semi-valent oxide.\cite{ACP92,ABRM99} 
It belongs to the homometallic warwickite family with formal chemical formula 
$MM'OBO_3$, where $M$ and $M'$ are, respectively, a divalent and trivalent metal 
ions. Surprisingly, that the homometallic ($M = M'$) warwickites are known only 
for Fe~\cite{ACP92,ABRM99} and Mn.\cite{GWA04,NKS95} In both compounds the metal 
have octahedral coordination. These octahedra share edges to form ribbons of four 
infinite along crystallographic $a$-direction chains of octahedra linked by corner 
sharing and the trigonal  BO$_3$ groups (see Fig.~\ref{fig:struc}). 

There are two crystallographically inequivalent sites of the metal ions Fe(1) 
and Fe(2). Fe$_2$OBO$_3$ is $L$-type ferrimagnetic with drastically smaller in 
comparison to Fe$_3$O$_4$ critical temperature of $T_c \approx 155$ K, the Fe(1)
magnetic moments being aligned antiparallel to the Fe(2) moments. It is almost 
antiferromagnetic but a small ferrimagnetic moment of $\sim 0.03$~$\mu_B$ per 
Fe atom in a 0.05 T field was found.\cite{ABRM99,ABRM98,ACP92,CPG01} 
At room temperature Fe$_2$OBO$_3$ is a semiconductor with a thermoactivated 
conductivity low $e^{-E_a/kT}$ with $E_a \approx 0.35$ eV.\cite{ABRM99,ABRM98} 
Upon farther heating a broad semiconductor-to-semiconductor transition occurs 
at $T_{co} \approx 317$ K, where resistivity drops down by a factor of
$\sim 3$, and, as a result, a small decreasing of the activated energy up to 
$E_a \approx 0.31$ eV above 350 K is observed.\cite{ABRM99, ABRM98} 
The 317 K transition is assigned to charge ordering of 2+ and 3+ Fe cations 
on Fe(1) and Fe(2) sites, and accompanied by a structural transition from 
monoclinic $P2_1/c$ to orthorhombic $Pmcn$ symmetry with increasing 
temperature. This structural transition is attributed by modification of the 
$\beta$ angle from $\beta = 90.220(1)^\circ$ at 3 K to $\beta = 90^\circ$ at 
337 K.\cite{ABRM98,ABRM99,ACP92} The change in conductivity and structure 
are small. But the $^{57}$Fe M$\ddot{\rm o}$ssbauer spectra at around 317 K 
clearly result in the charge localization at the transition with an equal
distribution of Fe$^{2+}$ and Fe$^{3+}$ cations over the two structurally 
distinct Fe(1) and Fe(2) sites with formal chemical formula 
Fe(1)$^{2+}_{0.5}$Fe(1)$^{3+}_{0.5}$Fe(2)$^{2+}_{0.5}$Fe(2)$^{3+}_{0.5}$OBO$_3$.\cite{ABRM99,
ABRM98,DPM00,ACP92,hiDPM00,CPG01,SKN03}
Although, there are two types of distorted FeO$_6$ octahedra with Fe-O bond length 
varying between 1.92 and 2.23 \AA~for 3 K, the average Fe(1)-O and Fe(2)-O distances are 
2.085 and 2.082 \AA, respectively, i.e. equal within experimental errors.\cite{ABRM99} 
Such a small difference results in the extremely small value of deviation ($\leq$ 0.01) 
from the average 2.5+ value of valence of Fe cations estimated by the bond valence sum 
method. While an electronic transition between charge ordered and disordered state occurs 
at around 317 K, as evidenced by M$\ddot{\rm o}$ssbauer spectroscopy and resistivity 
measurements, no long range Fe$^{2+}$/Fe$^{3+}$ ordering is directly observed by x-ray, 
neutron or electron diffraction. Thus, a long range charge ordering such as the simple 
alternating scheme proposed in Ref.~\onlinecite{ABRM98} destroys the mirror symmetry, 
which leads to a tilting of the Fe-ribbons, consistent with the observed enlargement 
of the $\beta$ angle below the transition. However, there is no observation of the 
increasing of $a$-axis periodicity (it should increase by a factor of two or another 
integer factor below $T_{co}$). Thus, below the transition, a charge ordering is not 
implicit in the atom coordinates, although it is indirectly evidenced by other 
experiments. This ambiguity is resolved in our electronic structure study, which 
reveals an arrangement of Fe$^{2+}$ and Fe$^{3+}$ cations alternately ordered within 
the chains along the $a$-direction.

In this paper we report theoretical investigation of the electronic structure 
and magnetic properties of Fe$_2$OBO$_3$ in the low-temperature $P2_1/c$ structure. 
The LSDA+$U$ approach in the tight-binding linear muffin-tin orbital (TB-LMTO) 
calculation scheme~\cite{AZA91,LAZ95} has been used. Motivated by our results, 
we propose an order parameter, defined as the difference between $t_{2g}$ 
minority spin occupancies of Fe(1)$^{2+}$ and Fe(1)$^{3+}$ as well as the 
difference between $t_{2g}$ majority spin occupancies of Fe(2)$^{2+}$ and 
Fe(2)$^{3+}$ cations. This order parameter is found to be quite large, 
although the total $3d$ charge difference between 2+ and 3+ cations, is small. 


\section{Computational details}
\label{sec:details}

The present band-structure calculations have been carried out for the 
low-temperature monoclinic structure of Fe$_2$OBO$_3$. The corresponding 
$P2_1/c$ unit cell contains four Fe$_2$OBO$_3$ formula units. We used 
atom coordinates from Ref.~\onlinecite{ABRM98} (see Table 1) and the cell 
parameters $a=3.1688$ \AA, $b=9.3835$ \AA, $c=9.2503$ \AA, and 
$\beta=90.22^\circ$ refined at 3 K.\cite{ABRM98, ABRM99, ACP92} 
The radii of muffin-tin spheres were taken $R_{\rm Fe}=2.5$ a.u., 
$R_{\rm B}=1.275$ a.u., $R_{\rm O1,O4}=1.7$ a.u., and $R_{\rm O2,O3}=1.8$ a.u.
Seven kinds of empty spheres were introduced to fill up the inter-atomic space.
For simplicity we neglect small spin-orbit coupling (for instance, in other iron 
oxide, in Fe$_3$O$_4$, the spin-orbital interaction for $3d$ electrons was found 
to be negligibly small). We consider only a collinear spin case, in which the Fe 
magnetic moments are aligned along the $a$-direction (see Table 1 of 
Ref.~\onlinecite{ACP92}), which is in a reasonably good agreement with 
the experimental magnetic structure of Fe$_2$OBO$_3$ below $T_c$.

\section{LSDA band structure}
\label{sec:lsda}

\begin{figure}[tbp!]
\centerline{\includegraphics[width=0.45\textwidth,clip]{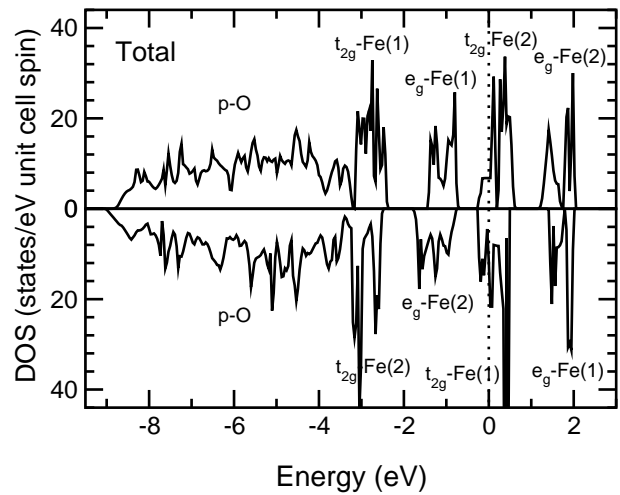}}
\caption{\label{fig:lsda}
Total density of states (DOS) obtained from the LSDA calculations for the 
low-temperature $P2_1/c$ phase of Fe$_2$OBO$_3$. The Fermi level  
is shown by dotted line.}
\end{figure}

The LSDA calculations give only a metallic ferrimagnetic solution 
without charge separation where partially filled bands at the Fermi 
level originate from the $t_{2g}$ orbitals of Fe cations 
(see Fig.~\ref{fig:lsda}). 

The lower part of the valence band (below -3.5 eV) is mainly formed by O $2p$ 
states with a bonding hybridization with Fe $3d$ states. Fe $3d$ states give 
predominant contribution to the bands at -3.5 eV below and up to 2.5 eV above 
the Fermi level. The exchange splitting between the spin-up and spin-down Fe 
$3d$ states is roughly 3 eV. 
Additionally, the five-fold $3d$ levels are 
split by the crystal field into $t_{2g}$ and $e_g$ sub-bands. The oxygen 
octahedra in Fe$_2$OBO$_3$ are strongly distorted and the local symmetry of 
Fe sites is, of cause, lower than cubic. Nevertheless, the cubic component 
of the ligand field, which is determined by the relative strength of 
Fe $d$ -- O $p$ hybridization of $\pi$- and $\sigma$-type, remains dominant, 
whereas the splitting within ``$t_{2g}$'' and ``$e_{g}$'' subbands is smaller 
than the corresponding band width. This allows one to label the corresponding
states as $t_{2g}$ and $e_{g}$. The crystal field splitting is roughly 2
eV, which is less than the exchange splitting. This is consistent with the
high-spin state of the Fe cations. The symmetry inequivalence of Fe(1) 
and Fe(2) sites leads to an inexact cancellation of magnetic moments and
results in a small ferrimagnetic moment of $\sim 0.31$~$\mu_B$ per formula
unit. The absolute values of magnetic moments obtained by LSDA are 3.54 
$\mu_B$ and 3.81 $\mu_B$ for Fe(1) and Fe(2) sites, respectively.

Fe(1) and Fe(2) $t_{2g}$ and $e_g$ states with the opposite spin
projections share nearly the same energy intervals. Thus, Fe $3d$ states
between -3.5 and -2.0 eV originate predominantly from majority spin Fe(1)
and minority spin Fe(2) $t_{2g}$ states whereas the states between -2.0 and
-0.5 eV are mainly of $e_g$ character. Partially occupied bands crossing
the Fermi level are formed by minority spin Fe(1) and majority spin Fe(2)
$t_{2g}$ states. The nominal occupation of these bands is 1/6. In the 
majority spin channel, however, the Fe(2) $t_{2g}$ state, that is oriented 
in the plane perpendicular to the shortest Fe(2)--O bond, forms 
quasi-one-dimensional bands with a strong dispersion along the $a$-direction. 
The one-dimensional character of the dispersion is explained by the fact that 
there are only two nearest neighbours of the same kind around each
Fe(2) ion. The other two Fe(2) $t_{2g}$ states are shifted to higher
energy and the corresponding bands are completely unoccupied. As a 
result, the majority spin bands crossing the Fermi level turn out to 
be half-filled. An Fe(1) ion, in contrast to Fe(2) one, has four Fe(1) 
neigbours at close distances. As a result of the hybridisation between 
Fe(1) $t_{2g}$ states the situation in the minority spin chanel is more 
complicated. Twelve $t_{2g}$ bands are split into three groups of 4 bands 
each. The Fermi level is crossed by lowest bands which show a rather 
strong dispersion along $a$-direction but with a period two times smaller
than the quasi-one-dimensional Fe(2) bands.

It should be noted that in contrast to experimental data~\cite{ABRM98, 
DPM00, hiDPM00} LSDA predicts Fe$_2$OBO$_3$ to be metallic with substantial 
magnetic moment per unit cell. Apparently, the electron-electron correlations, 
mainly in the $3d$ shell of Fe cations, play a significant role.


\section{LSDA+$U$ results and charge ordering}
\label{sec:co}

\begin{figure}[tbp!]
\centerline{\includegraphics[width=0.45\textwidth,clip]{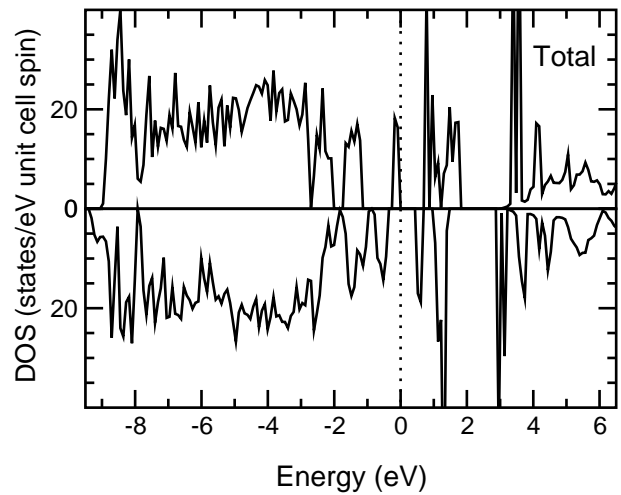}}
\caption{\label{fig:lsdau}
The total DOS obtained from LSDA+$U$ calculations with $U$=5.5 eV and $J$=1 eV  
for the low-temperature $P2_1/c$ phase of Fe$_2$OBO$_3$. The top of the valence 
band is shown by dotted lines.}
\end{figure}

To proceed further we take into account the strong electronic correlations 
in Fe $3d$ shell using the LSDA+$U$ method. The calculations have been 
performed for the $P2_1/c$ unit cell as well as for double 
($2a \times b \times c$) and triple ($3a \times b \times c$) $P2_1/c$ 
supercells of Fe$_2$OBO$_3$ (without putting in any additional local
displacements of oxygen atoms around of Fe$^{2+}$/Fe$^{3+}$ sites). 
For the $2a \times b \times c$ supercell the LSDA+$U$ calculations using
the classical value of Coulomb ($U$=5 eV) and exchange ($J$=1 eV)
interaction parameters for Fe result in an insulating charge ordered (CO)
solution with an energy gap of 0.13 eV. This is in a strong contrast to
metallic solution without CO obtained by the LSDA. The CO pattern obtained 
from the calculations is similar to the one proposed in Ref.~\onlinecite{ABRM98}
This is a notable result because 
the CO is not implicit in the atomic coordinates, and it shows that LSDA+$U$ 
calculations can assist experiments in revealing CO arrangements. To 
obtain a reasonably good agreement of the calculated gap of 0.39 eV with 
experimental value of 0.35 eV we increase the $U$ value up to 5.5 eV 
(see Fig.~\ref{fig:lsdau}). It does not exceed of 10\% of the $U$ value, 
which is in an accuracy of the $U$ calculation. Note, however, that the 
CO obtained by LSDA+$U$ within $2a \times b \times c$ supercell does not 
depend on the $U$ value of 5-5.5 eV. Here and in the following all results 
are presented for the double along $a$-direction $P2_1/c$ supercell of
Fe$_2$OBO$_3$. 

After self-consistency each of two groups of Fe(1) and Fe(2) atoms is 
split out in two subgroups of 2+ and 3+ Fe cations with equal number 
of 2+ and 3+ cations. Thus, one of $t_{2g}$ majority/minority spin 
states of Fe(2)/Fe(1) atom becomes completely occupied, whereas all 
other $t_{2g}$ states are pushed by strong Coulomb interaction at 
the energies above 3 eV. The gap is opened between occupied and 
unoccupied $t_{2g}$ states of Fe(1)$^{2+}$ and Fe(1)$^{3+}$ for 
spin-down and Fe(2)$^{2+}$ and Fe(2)$^{3+}$ for spin-up. Majority 
spin $3d$ states of Fe(1)$^{3+}$ and minority spin states of 
Fe(2)$^{3+}$ cations are shifted below the O $2p$ states, which
form the band in the energy range of -8 and -2 eV. In contrast to 
Fe$^{3+}$ states, the majority spin Fe(1)$^{2+}$ and minority spin 
Fe(2)$^{2+}$ $3d$ states form the broad bands between -8 and -1 eV. 

\begin{figure}[tbp!]
\centerline{\includegraphics[width=0.45\textwidth,clip]{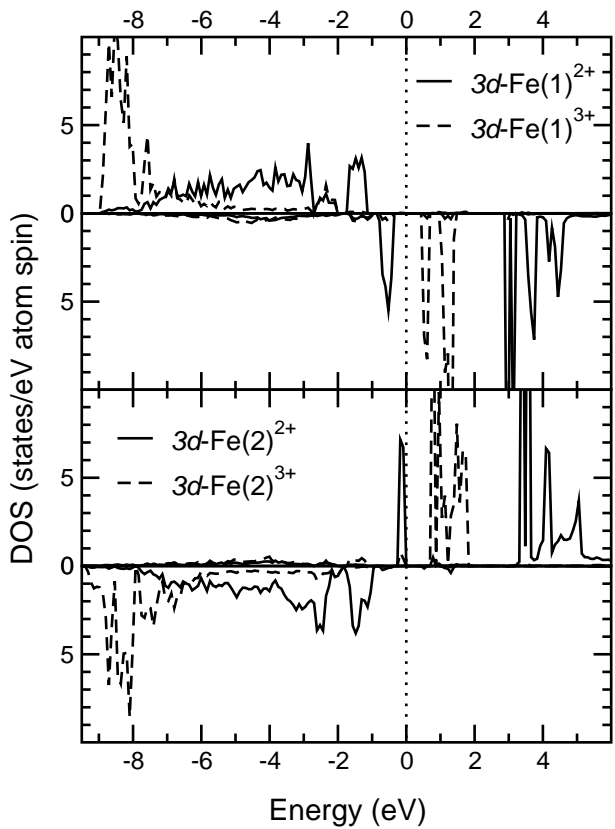}}
\caption{\label{fig:lsdau-pdos}
The partial DOS for different Fe cations are shown. The gap is opened 
between Fe(2)$^{2+}$ and Fe(2)$^{3+}$ for majority spin and Fe(1)$^{2+}$ 
and Fe(1)$^{3+}$ cations for minority spin states. The gap value of 
0.39 eV was obtained by LSDA+$U$ with $U$=5.5 eV and $J$=1 eV.
The Fermi level is shown by dotted line.}
\end{figure}


The obtained magnetic structure is almost antiferromagnetic (without spin moment 
per unit cell) with nearly the same spin moment per Fe(1)$^{2+}$ and Fe(2)$^{2+}$ 
as well as per Fe(1)$^{3+}$ and Fe(2)$^{3+}$ cations. Using the moment populations 
in Table~\ref{tab:occ}, the calculated net moment is $\sim 0.04~\mu_B$ per Fe atom, 
in exact agreement with the experimental value (Ref.~\onlinecite{ABRM98}).


The charge order obtained by LSDA+$U$ in $2a\times b\times c$ $P2_1/c$ supercell
is consistent with observed enlargement of the $\beta$ angle below the transition 
and coincides with charge ordering scheme proposed earlier by 
Attfield $et~al.$ in \onlinecite{ABRM98}. It is described by the sloping 2+ 
and 3+ Fe cation 
lines alternately stacked along $a$-direction, and could be considered as 
a quasi one dimensional analog of the Verwey CO model in the pyrochlore lattice 
of Fe$_3$O$_4$. Using the same $U$ and $J$ values we perform additional 
self-consistent LSDA+$U$ calculations for $P2_1/c$ unit cell as well as for 
double and triple along $a$-direction $P2_1/c$ supercells. But only 
self-consistent solutions with larger value of the total energy or
with substantial magnetic moment per unit cell, which contradicts 
to the experimental data, were found. Also we found that other charge 
arrangements in $2a\times b \times c$ $P2_1/c$ supercell are unstable, 
and the stable one coincides with the CO found previously. Thus, the CO 
obtained for certain value of $U$ and $J$ does not depend on the initial 
charge arrangemet. It is not possible to check all possible CO arrangements 
including more complex CO scenarios, but our results consistently indicate 
that the obtained CO solution is more favourable than other simple alternatives, 
and is the ground state of Fe$_2$OBO$_3$ in the low-temperature phase.

Because of the small monoclinic distortion of the low temperature $P2_1/c$
structure the distances between the nearest Fe(1) cations become slightly
different. It is important to note that in the calculated CO pattern the
shortest (2.957 \AA) is the distance between the pairs of equally charged
Fe(1) cations whereas the distance between Fe(1)$^{2+}$ and Fe(1)$^{3+}$
cations turns out to be larger (2.961 \AA). On the contrary, the pairs of
Fe(1) and Fe(2) sites with the shortest distance between them are occupied
by the cations of different valency. Thus, assuming a simple ionic model
with the Coulomb interaction between Fe ions only the obtained CO pattern
contradicts to the requirement of the minimal electrostatic energy. Indeed,
the comparison of the Madelung energies shows that the energy lowers 
(about 0.07 eV per unit cell) if the nearest Fe(1) sites are occupied by 
Fe(1)$^{2+}$ and Fe(1)$^{3+}$ ions. This observation suggests that 
electron-lattice coupling rather than electrostatic repulsions drives 
the charge ordering in Fe$_2$OBO$_3$; the same conclusion was found for the 
charge order in Fe$_3$O$_4$.\cite{WAR01,WAR02,LYA04}


Although the corresponding total $3d$ charges difference $(0.34\bar{e})$ and 
disproportion of the total electron charges inside the atomic spheres 
of Fe$^{2+}$ and Fe$^{3+}$ cations $(0.24\bar{e})$ is small, an analysis of 
occupation matrices of $3d$ Fe(1)/Fe(2) minority/majority spin 
states confirms substantial charge separation. Thus,
as shown in Table~\ref{tab:occ}, one of the $t_{2g}$ states of
Fe(1)$^{2+}$ and Fe(2)$^{2+}$ cations is almost completely
filled with the occupation numbers $n\approx 0.9$, whereas 
the remained two $t_{2g}$ orbitals of the Fe$^{2+}$ cations have 
significantly smaller population of about 0.1. 
According to Ref.~\onlinecite{LYA04} we define an order parameter as 
the largest difference 
between Fe$^{2+}$ and Fe$^{3+}$ $t_{2g}$ populations. 
While due to strong static ``screening'' effects,\cite{screening} the order 
parameter introduced as the total $3d$ charge difference between 2+ and 3+ 
Fe cations is ill defined, the well-defined order parameter is the difference 
of $t_{2g}$ occupancies for Fe$^{3+}$ and Fe$^{2+}$ cations, which amounts 
to 80\% of ideal ionic CO model and clearly pronounces the existence of CO 
below the transition. The occupation matrices analysis shows that the 
change of the $t_{2g}$ occupations caused by the charge ordering is very 
effectively screened by the rearrangement of the other Fe electrons. Thus, 
significant contribution to the charge screening is provided by Fe $e_g$ 
states due to relatively strong $\sigma$ bonds with 2$p$ O states and, as 
a result, appreciable contribution to the occupied part of the valence band.

\begin{table}[tbp!]
\caption{\label{tab:occ}Total and $l$-projected charges, magnetic moments, and
occupation of the most populated $t_{2g}$ orbitals calculated for
inequivalent Fe atoms in the low-temperature $P2_1/c$ phase of Fe$_2$OBO$_3$.}
\begin{ruledtabular}
\begin{tabular}{lccccccc}
Fe ion & $q$ &$q_s$ &$q_p$ &$q_d$ & $M$ ($\mu_{\text{B}}$) & $t_{2g}$ orbital & $n$ \\
\hline
Fe(1)$^{3+}$  & 6.90 & 0.40 & 0.55 & 5.95 & -4.20 & - & 0.10 \\
Fe(1)$^{2+}$  & 7.12 & 0.35 & 0.50 & 6.27 & -3.65 & $d_{xy\downarrow}$ & 0.91 \\
Fe(2)$^{3+}$  & 6.79 & 0.38 & 0.54 & 5.86 &  ~4.33 & - & 0.09 \\
Fe(2)$^{2+}$  & 7.04 & 0.34 & 0.50 & 6.21 &  ~3.69 & $d_{xy\uparrow}$ & 0.89 \\
\end{tabular}
\end{ruledtabular}
\end{table}

\begin{figure}[tbp!]
\centerline{\includegraphics[width=0.45\textwidth,clip]{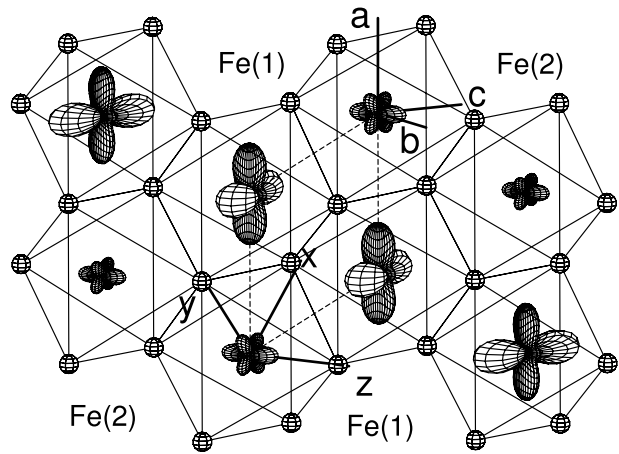}}
\caption{\label{fig:orb} 
The angular distribution of the majority and minority spin $3d$ electron 
density of the Fe(2) and Fe(1) cations, respectively, within Fe-ribbon.
The size of orbital corresponds to its occupancy. Oxygen atoms are shown 
by small spheres. X-Y-Z coordinate system corresponds to the local cubic 
frame.}
\end{figure}
\begin{figure}[tbp!]
\centerline{\includegraphics[height=0.3\textheight,clip]{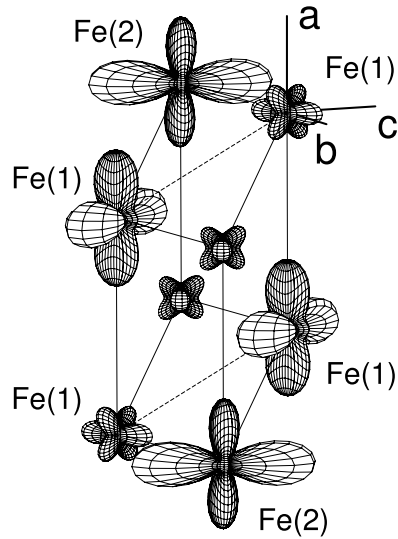}}
\caption{\label{fig:inter_orb} 
The angular distribution of the majority and minority spin $3d$ electron 
density of the Fe(2) and Fe(1) cations, respectively, from different 
Fe-ribbons.\cite{chargedens} The size of orbital corresponds to its 
occupancy. The frame of four Fe(1) atoms from the Fe-ribbon presented 
in Fig.~\ref{fig:orb} is shown by dashed lines.}
\end{figure}

The occupied $t_{2g}$ states of Fe$^{2+}$ cations are predominantly of $d_{xy}$ 
character in the local cubic frame (according to that we later mark the orbital as
$d_{xy}$ orbital). This is illustrated in Fig.~\ref{fig:orb}, which shows the 
angular distribution of the majority and minority spin $3d$ electron density 
of the Fe(2) and Fe(1) cations, respectively.\cite{chargedens} Thus, occupied 
Fe$^{2+}$ and unoccupied Fe$^{3+}$ cations are ordered alternately within the 
chain which is infinite along $a$-direction. The angular distribution of charge 
density of the Fe(1) and Fe(2) cations, which correspondingly belongs to different 
Fe-ribbons being formed a cross in the Fe$_2$OBO$_3$ structure projected on (100) 
plane (see Fig.~\ref{fig:struc}) is shown in Fig.~\ref{fig:inter_orb}.
An analysis of interatomic distances (Table~\ref{tab:dist}) shows that the 
average Fe(2)-O distance (2.109 \AA) in the plane perpendicular to one of the 
diagonals of the distorted Fe(2)O$_6$ octahedron is considerably larger than 
average distances in the other two planes (2.055 and 2.083 \AA). It turns out 
that the occupied Fe(2) $t_{2g}$ majority spin orbital is the one oriented in 
the plane with the largest average Fe(2)-O distance. The same is also true for 
the Fe(1) ion but in this case the variation of the average Fe(1)-O distances 
is smaller (2.111 vs 2.069 and 2.076 \AA) and, as a consequence, the out-of-plane 
rotation of the occupied $t_{2g}$ minority spin orbital is stronger.

\begin{table}[tbp!]
\caption{\label{tab:dist}The averaged Fe--O distances in the plane of 
$t_{2g}$ orbitals for $P2_1/c$ structure of Fe$_2$OBO$_3$. $d_{xy}$ 
approximates to the doubly-occupied orbital of the $3d^6$ Fe$^{2+}$ states.}
\begin{ruledtabular}
\begin{tabular}{lccc}
Fe atom & orbital & $d_{\mathsf{orb.}}$ (\AA) & $d_{\mathsf{av.}}$ (\AA)\\
\hline
Fe(1)                 & $d_{xy}$   & 2.111 & 2.085  \\
                      & $d_{yz}$   & 2.076 &        \\
                      & $d_{zx}$   & 2.069 &        \\
Fe(2)                 & $d_{xy}$   & 2.109 & 2.082  \\
                      & $d_{yz}$   & 2.083 &        \\
                      & $d_{zx}$   & 2.055 &        \\
\end{tabular}
\end{ruledtabular}
\end{table}


\section{Exchange coupling constants}
\label{sec:j}

\begin{figure}[tbp!]
\centerline{\includegraphics[width=0.425\textwidth,clip]{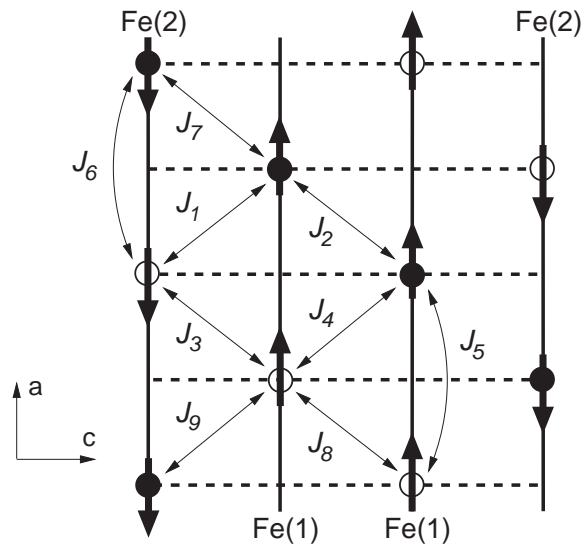}}
\caption{\label{fig:intra_exchange} 
The sketch of the arrangement of exchange interaction parameters within the 
ribbon of iron atoms. Open circles correspond to 
Fe$^{3+}$, while Fe$^{2+}$ cations are noted by the closed circles. The spin 
moment direction on each Fe site is shown by an arrow.}
\end{figure}

Using the LSDA+$U$ method the exchange interaction parameters have been calculated 
via the variation of ground state energy with respect to the magnetic-moment rotation 
angle.\cite{LAZ95,AAL97} In Table~\ref{tab:intra_exchange} we have shown the total 
set of different intraribbon exchange parameters as well as a contribution of 
different subbands into exchange interactions. $J_i$ represents the effective pair 
exchange interaction between Fe atoms with effective Heisenberg Hamiltonian 
$H = - \sum_{i > j} J_{ij} e_i \cdot e_j$, where $e_i$ and $e_j$ are magnetic 
moment unit vectors at site $i$ and $j$. Positive (negative) values of $J$ 
correspond to the ferromagnetic (antiferromagnetic) coupling between Fe sites.
The spatial representation of all 
these exchanges is schematically presented in Fig.~\ref{fig:intra_exchange}. 
Surprisingly, only the exchange interaction parameter between Fe(2)$^{2+}$ and 
Fe(2)$^{3+}$ cations is ferromagnetic with relatively small value of $J_6 = 6$ K. 
In contrast, the nearest sites in quasi-one-dimensional Fe(1) chain are coupled 
antiferromagnetically with noticeably larger exchange absolute value of $|J_5| = 69$ K. 
Furtermore, the exchange parameters between the nearest sites of two Fe(1) chains are 
relatively strong and antiferromagnetic (see $J_2$, $J_4$, and $J_8$ in 
Table~\ref{tab:intra_exchange}). Therefore, the Fe(1) sublattice is highly frustrated, 
while the relatively weak frustrations in the Fe(2) sublattice considerably reduce 
ferromagnetic interaction within Fe(2) chain. Also it is interesting to note that 
relatively strong ferromagnetic intrachain interaction between $t_{2g}$ subbands 
of Fe(2)$^{2+}$ and Fe(2)$^{3+}$ cations (see $J_6$ in Table~\ref{tab:intra_exchange}) 
is strongly suppressed by the substantial antiferromagnetic $t_{2g}$ -- $e_g$ and 
$e_g$ -- $e_g$ exchange.

\begin{table}[tbp!]
\caption{\label{tab:intra_exchange}
Total and partial intraribbon exchange interaction 
parameters are shown. The values are given in kelvin.
The spatial representation of all these exchanges is schematically presented 
in Fig.~\ref{fig:intra_exchange}.}
\begin{ruledtabular}
\begin{tabular}{lccccc}
 $J_i$ & $t_{2g}-t_{2g}$ & $t_{2g}-e_g$ & $e_g-e_g$ & Total \\
\hline
$J_1$  & ~4  & -22  & -20 & -38 \\
$J_2$  & -7  & -38  & ~1  & -44 \\
$J_3$  & ~3  & -19  & -16 & -32 \\
$J_4$  & ~4  & -72  & -1  & -70 \\
$J_5$  &  17 & -69  & -16 & -69 \\
$J_6$  &  53 & -34  & -13 & ~6  \\
$J_7$  & -1  & -11  & -4  & -16  \\
$J_8$  & -13 & -120 & -4  & -136 \\
$J_9$  & ~3  & -22  & -20 & -39 \\
\end{tabular}
\end{ruledtabular}
\end{table}

\begin{table}[tbp!]
\caption{\label{tab:inter_exchange}
Total and partial interribbon exchange interaction 
parameters are shown. The values are given in kelvin.
The spatial representation of all these exchanges is schematically 
presented in Fig.~\ref{fig:inter_exchange}.}
\begin{ruledtabular}
\begin{tabular}{lccccc}
 $J_i$ & $t_{2g}-t_{2g}$ & $t_{2g}-e_g$ & $e_g-e_g$ & Total \\
\hline
$J_{10}$  & -12 & -114 & -104 & -229 \\
$J_{11}$  & -16 & -140 & -28  & -183 \\
$J_{12}$  & -17 & -141 & -52  & -209 \\
$J_{13}$  & -3  & -52  & -2   & -56  \\
$J_{14}$  & -22 & -165 & -20  & -207 \\
$J_{15}$  & -12 & -115 & -19  & -147 \\
\end{tabular}
\end{ruledtabular}
\end{table}

\begin{figure}[tbp!]
\centerline{\includegraphics[width=0.375\textwidth,clip]{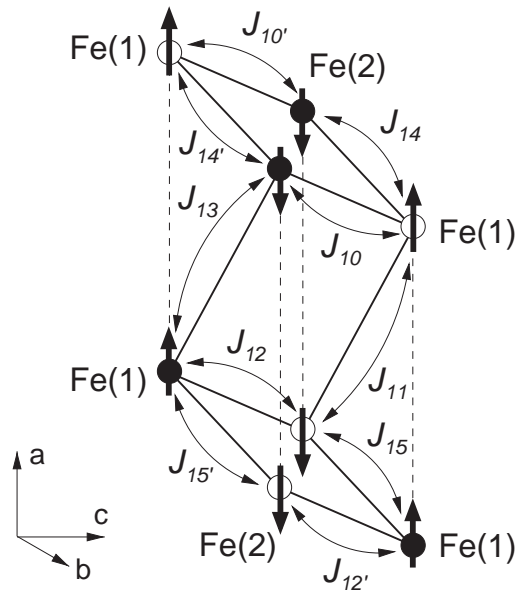}}
\caption{\label{fig:inter_exchange} 
The sketch of the arrangement of interribbon exchange interaction parameters. 
Open circles correspond to Fe$^{3+}$, while Fe$^{2+}$ cations are noted by the 
closed circles. The spin moment direction on each Fe site is shown by arrow. 
Note that $J_{i'}$ exchange parameters presented here have the same total 
values as $J_{i}$, while subband contributions are different.}

\end{figure}

On the other hand, the interribbon exchange interaction parameters between Fe(1) 
and Fe(2) atoms are considerably larger. The values of these interactions 
are shown in Table~\ref{tab:inter_exchange}, whereas the spatial representation is 
schematically presented in Fig.~\ref{fig:inter_exchange}. Thus, the exchange parameters
between Fe(1)$^{3+}$ and Fe(2)$^{2+}$ cations are antiferromagnetic with values of 
$J_{10} =$-229 K and $J_{14}=$-207 K (see Table~\ref{tab:inter_exchange}). 
Such an appreciable difference between $J_{10}$ and $J_{14}$ arises from
geometry since the former is due to the superexchange interaction between
Fe ions linked by shortest bonds to a common O ion.
It seems that the geometrical reason is also responsible for decreasing of absolute 
value of the exchange interactions between Fe(1)$^{2+}$ and Fe(2)$^{3+}$ cations from
$|J_{12}| = 209$ K to $|J_{15}| = 147$ K (see Table~\ref{tab:inter_exchange}).
Also it is interesting to note that the exchange interaction between Fe(1)$^{3+}$ and
Fe(2)$^{3+}$ cations  is considerably larger than between Fe(1)$^{2+}$ and 
Fe(2)$^{2+}$ ($J_{11}$ and $J_{13}$, respectively). We find that the interribbon 
exchange interactions play predominant role and determine the whole $L$-type 
ferrimagnetic spin structure below $T_c$ in contrast to the ferromagnetic intrachain 
order due to $d^5$ -- $d^6$ superexchange.\cite{ABRM98,ABRM99}


\section{Summary and conclusions}
\label{sec:sum}

In summary, in the present LSDA+$U$ study of the low-temperature $P2_1/c$ phase 
of Fe$_2$OBO$_3$ we found a charge ordered insulator with an energy gap of 0.39 eV. 
While the screening of the charge disproportion is so effective that the total
$3d$ charge separation is rather small (0.34), the charge order is well
pronounced with an order parameter defined as a difference of $t_{2g}$ occupancies 
of 2+ and 3+ Fe cations (0.8). The occupied Fe$^{2+}$ and Fe$^{3+}$ cations are 
ordered alternately within infinite along $a$-axis chains of Fe atoms.
This result is remarkable in view of the absence of directly observed CO atomic
displacements in the experimental coordinates, and demonstrates the utility of
the LSDA+$U$ method as an aide to experimental studies of CO structures.
However, the charge order obtained by LSDA+$U$ is consistent with observed 
enlargement of the $\beta$ angle and coincides with charge ordering scheme 
proposed earlier by Attfield $et~al.$\cite{ABRM98} It seems certain that 
Fe$_2$OBO$_3$ is charge ordered below $T_{co}$, and the absence of the long 
range charge ordering from x-ray, neutron or electron diffraction arises 
from formation of charge order within small domains, which have been 
termed ``Wigner nanocrystals''.\cite{RHP96} Thus, the superstructure 
peaks are too weak and broad to be observed against background in 
diffraction patterns, whereas the observed long range monoclinic lattice 
distortion can arise despite a large concentration of defects as these 
preserve the direction of the monoclinic distortion, but do not propagate 
the coherent doubling of the lattice periodicity. An analysis of the 
exchange interaction parameters obtained by LSDA+$U$ method inevitably 
results in predominance of the interribbon exchange interactions which 
determine the whole $L$-type ferrimagnetic spin structure below $T_c$, 
in contrast to the ferromagnetic intrachain order due to $d^5$ -- $d^6$ 
superexchange proposed earlier in Ref.~\onlinecite{ABRM98}.

We are grateful to M.~A.~Korotin, R.~Claessen, P.~Fulde, 
and D.~Vollhardt for helpful discussions. The present work was supported in part by 
RFFI Grant No. 04-02-16096, No. 03-02-39024, and by the Sonderforschungsbereich 484 
of the Deutsche Forschungsgemeinschaft (DFG).


\begin{thebibliography}{20}
\bibitem{screening} 
Here and in the following we assume a redistribution of charge between Fe 
$t_{2g}$ and other states using term screening.
\bibitem{chargedens} 
The distribution is calculated according to 
$\rho(\theta,\phi)=\sum_{m,m'}n_{m,m'}~Y^\ast_{m}(\theta,\phi)~Y_{m'}(\theta,\phi)$,
where $n_{m,m'}$ is the occupation matrix of $3d$ minority states of Fe(1) and
$3d$ majority states of Fe(2) cations. The occupation matrices were calculated by 
the LSDA+$U$ with $U$= 5.5 eV, $J$= 1 eV for the low-temperature $P2_1/c$ phase 
of Fe$_2$OBO$_3$. $Y_{m}(\theta,\phi)$ are corresponding spherical harmonics.
\end{thebibliography}
\end{document}